\documentclass[12pt]{article}
\usepackage{amsmath}
\usepackage{amsfonts}

\def\gtwid{\mathrel{\raise.3ex\hbox{$>$\kern-.75em\lower1ex\hbox{$\sim$}}}}
\def\ltwid{\mathrel{\raise.3ex\hbox{$<$\kern-.75em\lower1ex\hbox{$\sim$}}}}
\def\square{\kern1pt\vbox{\hrule height 1.2pt\hbox{\vrule width 1.2pt\hskip 3pt
   \vbox{\vskip 6pt}\hskip 3pt\vrule width 0.6pt}\hrule height 0.6pt}\kern1pt}

\begin{document}

\begin{titlepage}

\begin{flushright}
UFIFT-QG-17-02 \\
\end{flushright}

\vskip .5cm

\begin{center}
{\bf How to Produce an Arbitrarily Small Tensor to Scalar Ratio}
\end{center}

\vskip 1cm

\begin{center}
D. J. Brooker$^{1*}$
\end{center}

\vskip .5cm

\begin{center}
\it{$^{1}$ Department of Physics, University of Florida,\\
Gainesville, FL 32611, UNITED STATES}
\end{center}

\vspace{.5cm}

\begin{center}
ABSTRACT
\end{center}

We construct a toy a model which demonstrates that large field single scalar inflation can produce an arbitrarily small scalar to tensor ratio in the window of e-foldings recoverable from CMB experiments. This is done by generalizing the $\alpha$-attractor models to allow the potential to approach a constant as rapidly as we desire for super-planckian field values. This implies that a non-detection of $r$ alone can never rule out entirely the theory of large field inflation. 

\vskip .5cm

\begin{flushleft}
$^{*}$ e-mail: djbrooker@ufl.edu \\
\end{flushleft}

\end{titlepage}
\begin{section}{Introduction}
	The theory of inflation is of tremendous importance to modern cosmology as a means of solving the horizon, flatness and monopole problems with a single theory \cite{brout,starobinsky80,kazanas,sato,guth,linde82,Albrecht,linde83}. In addition, inflation gives wonderful predictions for the spectra of primordial perturbations \cite{Starobinsky:1979ty,mukhanov81}. As of yet however the prediction of a primordial tensor spectrum has eluded observational efforts \cite{Ade:2015lrj}. This opens up the question of whether or not large field inflation can accommodate an arbitrarily small tensor to scalar ratio. Previous attempts to answer this question in the context of large field inflation were restricted only to the $\alpha$-attractor models and found a lower bound of $r\simeq 2\times10^{-5}$ \cite{Garcia-Bellido:2014wfa,Linde:2016hbb}. We should note that in the context of small field inflation it has been known for some time that one can achieve arbitrarily small values of $r$ \cite{Kinney:1995cc,Kinney:1995ki}. In this work we will construct a toy model for which the tensor to scalar ratio seemingly has no lower bound. For the purpose of this exploration we will use a definition due to Linde that large field inflation corresponds to any theory in which the canonically normalized inflaton $\phi$ takes on a value greater than $M_p$ during the last 50-60 e-foldings of inflation \cite{Linde:2016hbb}. We will begin by first reminding ourselves of some basic properties of the $\alpha$-attractor models then we will examine our toy model which is a generalization of the $\alpha$-attractors. There are several ways to formulate the family of $\alpha$-attractors but for our purposes we will stick to the following potentials which generalize the Starobinsky model:
\begin{equation}
V \left( \phi \right) = V_0 \left( 1 - e^{-\sqrt{\frac{2}{3\alpha}}\phi}\right)^2 \, .
\end{equation}
In the slow roll approximation the number of e-foldings until the end of inflation can be expressed as an integral over $\phi$. That is:
\begin{equation}
N\simeq \int^{\phi _N}_{\phi _f} \frac{V(\phi)}{V'(\phi)} d\phi
\end{equation}
where the prime above indicates a derivative with respect to $\phi$. For the potentials in (1) we find that $N(\phi)$ can be given to very good approximation by:
\begin{equation}
N\simeq \frac{3\alpha}{4}e^{\sqrt{\frac{2}{3\alpha}}\phi_N} \longrightarrow \phi_N \simeq \sqrt{\frac{3\alpha}{2}}\log\left[\frac{4N}{3\alpha}\right] \, .
\end{equation}
The tensor to scalar ratio is related to the potential in the context of slow roll by:
\begin{equation}
r\simeq 8\left(\frac{V'(\phi)}{V(\phi)}\right)^2 \, .
\end{equation}
Inserting equation (3) into this yields the well known result:
\begin{equation}
r\simeq \frac{12\alpha}{N^2} \, .
\end{equation}
Since we require for our definition of large field inflation that $\phi > 1$ somewhere in the range $50<N<60$ we cannot achieve arbitrarily low values of $r$ simply by tuning $\alpha$. The reason for this is that for small enough $\alpha$ we come to a regime where the part of inflation accessible by the CMB is entirely sub-planckian. 
\end{section}
\begin{section}{Toy Model}
Our generalization will be to allow the factor of $\phi$ to be raised to an arbitrary power $n$. Our new potential is:
\begin{equation}
V \left( \phi \right) = V_0 \left( 1 - e^{-\sqrt{\frac{2}{3\alpha}}\phi^n}\right)^2 \, .
\end{equation}
We have explored this potential numerically for several values of $n$ and $\alpha$. The tensor to scalar ratio was evaluated by solving the full mode equations for both scalar and tensor fluctuations numerically using the methods outlined in references \cite{brookerT,brookerS}. To check consistency with current observations, the scalar spectral index ($n_S$) was evaluated using the first four terms in the slow roll expansion. Our results are summarized in table 1 with $r$ and $\phi$ being evaluated at $N=55$.
\begin{table}[h]
\begin{center}
\begin{tabular}{ccccc}
$\alpha$&$n$&$\phi_{55}$&$r_{55}$&$n_S$\\
\hline\\
$1$&$2$&$3.07$&$1.49\times10^{-4}$&$0.9591$\\
$0.1$&$2$&$1.87$&$3.8\times10^{-5}$&$0.9598$\\
$0.01$&$2$&$1.13$&$1.01\times10^{-5}$&$0.9603$\\
$1$&$4$&$1.90$&$7.62\times10^{-6}$&$0.9587$\\
$0.1$&$4$&$1.45$&$3.81\times10^{-6}$&$0.9591$\\
$0.01$&$4$&$1.10$&$1.92\times10^{-6}$&$0.9594$\\
$1$&$6$&$1.57$&$1.76\times10^{-6}$&$0.9590$\\
$0.1$&$6$&$1.30$&$1.11\times10^{-6}$&$0.9592$\\
$0.01$&$6$&$1.08$&$7.04\times10^{-7}$&$0.9593$\\
\hline\\
\end{tabular}
\caption{Values of $\phi$, $r$, and $n_S$ evaluated 55 e-foldings until the end of inflation for several generalized $\alpha$-attractors.}
\end{center}
\end{table}

The first thing about the table to note are the values of $n_S$ which tell us right away whether or not a theory is viable. The Planck constraints on inflation give
\begin{equation}
n_S=0.9677\pm 0.006 \;\left(68\% \, CL \right)
\end{equation}
as a value for the spectral index \cite{Ade:2015lrj}. Since all of the values for $n_S$ in table 1 are within the $95\%$ confidence interval of the most recent Planck data these models cannot be ruled on the grounds of current observations. We can understand why it is possible to separate the smallness of $r$ from the constraints imposed by Planck simply by looking at the first two non-trivial terms in the slow roll expansion for $n_S$. Those terms are:
\begin{equation}
n_s=1-2\epsilon-\frac{\partial_n\epsilon}{\epsilon} \; .
\end{equation}
It is easy to see that even in the limit where $\epsilon$ is very small we can still have any value of $n_S$ we like by choosing a functional dependence on $n$ such that the ratio $\partial_n\epsilon/\epsilon$ is as big or small as we desire. 

\paragraph{}
Now that we have seen that these models are consistent with the data we will examine the behavior of the tensor to scalar ratio. As can be seen from table 1 it would appear as though we can achieve any small value of $r$ we desire by doing two things. The first is that we make $\alpha$ small to reduce $r$ but not so small that $\phi$ be sub-planckian. Then we increase $n$ as much as we wish. We can understand this behavior better by looking at a rough analytic estimate for $r$. Using (2) we have that the number of e-foldings is given by:
\begin{equation}
N\simeq \sqrt{\frac{3\alpha}{8}}\frac{1}{n}\int^{\phi_N}\frac{e^{\sqrt{\frac2{3\alpha}}\phi^n}}{\phi^{n-1}}
\end{equation}
This integral can be solved approximately by assuming that most of the dependence on $\phi$ comes from the exponential. This tells us that:
\begin{equation}
N(\phi)\sim\frac{3\alpha}{4n^2}\phi_N^{2-2n}e^{a\phi_N^n}
\end{equation}
Then since our model is tuned so that $\phi_N$ will be nearly unity for modes relevant to CMB observations we can arrive at:
\begin{equation}
r\sim \frac{12\alpha}{N^2n^2}
\end{equation}
This dependence on both $\alpha$ and $n$ strongly suggests that we could reach \emph{any} arbitrarily small value of $r$ from a large field model of inflation. It is very important to note that we did not require any specific form of the potential to achieve such small values of $r$. All that was necessary was a free parameter, $n$ in this case, which can make $V'(\phi)$ as small as we like without pushing the field to sub-planckian values. In principle there may be many such potentials which can achieve these small values. 
\end{section}
\begin{section}{Conclusion}
We have shown that by modifying the form of the $\alpha$-attractor potential that there exist large field inflationary potentials which can achieve an arbitrarily small value of tensor to scalar ratio. We will close by discussing what this means for the theory of inflation as a whole. As a test of scalar driven inflation it would seem that attempts to measure the tensor to scalar ratio are in a sense one sided. On the one hand a detection of the primordial tensor spectrum would set the scale of inflation, allow us to reconstruct the inflation potential, and allow us to test the consistency of single scalar inflation. On the other hand it is apparent that a continued non-detection of primordial tensor fluctuations cannot on its own rule out large field single scalar inflation. This begs us to focus more theoretical effort on constructing other means of testing inflation such as novel probes of primordial non-gaussianity and reheating.
\end{section}
\begin{section}{Acknowledgments}
This work was partially supported by by NSF grant PHY1506513; and by the Institute for Fundamental Theory at the University of Florida. The author would like to thank Richard Woodard for his encouragement and helpful discussion.

\end{section}


\begin{thebibliography}{99}

\bibitem{brout}
R. Brout, F. Englert and E. Gunzig,
Annals Phys. \textbf{115}, 78 (1978).
doi:10.1016/0003-4916(78)90176-8

\bibitem{starobinsky80}
A.~A.~Starobinsky,
Phys. Lett. B \textbf{91}, 99 (1980).
doi:10.1016/0370-2693(80)90670-X

\bibitem{kazanas}
D. Kazanas,
Astrophys. J. \textbf{241}, L59 (1980). doi:10.1086/183361

\bibitem{sato}
K. Sato, Mon. Not. Roy. Astron. Soc. \textbf{195}, 467 (1981).

\bibitem{guth}
A. H. Guth, Phys. Rev. D \textbf{23}, 347 (1981). doi:10.1103/PhysRevD.23.347

\bibitem{linde82}
A. D. Linde, Phys. Lett. B \textbf{108}, 389 (1982). doi:10.1016/0370-
2693(82)91219-9

\bibitem{Albrecht}
A. Albrecht and P. J. Steinhardt, Phys. Rev. Lett. \textbf{48}, 1220 (1982).
doi:10.1103/PhysRevLett.48.1220

\bibitem{linde83}
A. D. Linde, Phys. Lett. B \textbf{129}, 177 (1983). doi:10.1016/0370-
2693(83)90837-7

\bibitem{Starobinsky:1979ty} 
  A.~A.~Starobinsky,
  JETP Lett.\  {\bf 30}, 682 (1979)
  [Pisma Zh.\ Eksp.\ Teor.\ Fiz.\  {\bf 30}, 719 (1979)].

\bibitem{mukhanov81}
V. F. Mukhanov and G. V. Chibisov, JETP Lett. \textbf{33}, 532 (1981) [Pisma
Zh. Eksp. Teor. Fiz. 33, 549 (1981)].
  
\bibitem{Ade:2015lrj} 
  P.~A.~R.~Ade {\it et al.} [Planck Collaboration],
  Astron.\ Astrophys.\  {\bf 594}, A20 (2016)
  doi:10.1051/0004-6361/201525898
  [arXiv:1502.02114 [astro-ph.CO]].

\bibitem{Garcia-Bellido:2014wfa} 
  J.~Garcia-Bellido, D.~Roest, M.~Scalisi and I.~Zavala,
  Phys.\ Rev.\ D {\bf 90}, no. 12, 123539 (2014)
  doi:10.1103/PhysRevD.90.123539
  [arXiv:1408.6839 [hep-th]].

\bibitem{Linde:2016hbb} 
  A.~Linde,
  JCAP {\bf 1702}, no. 02, 006 (2017)
  doi:10.1088/1475-7516/2017/02/006
  [arXiv:1612.00020 [astro-ph.CO]].
  
\bibitem{Kinney:1995cc} 
  W.~H.~Kinney and K.~T.~Mahanthappa,
  Phys.\ Rev.\ D {\bf 53}, 5455 (1996)
  doi:10.1103/PhysRevD.53.5455
  [hep-ph/9512241].
  
\bibitem{Kinney:1995ki} 
  W.~H.~Kinney and K.~T.~Mahanthappa,
  Phys.\ Lett.\ B {\bf 383}, 24 (1996)
  doi:10.1016/0370-2693(96)00713-7
  [hep-ph/9511460].

\bibitem{brookerT} 
  D.~J.~Brooker, N.~C.~Tsamis and R.~P.~Woodard,
  Phys.\ Rev.\ D {\bf 93}, no. 4, 043503 (2016)
  doi:10.1103/PhysRevD.93.043503
  [arXiv:1507.07452 [astro-ph.CO]].

\bibitem{brookerS} 
  D.~J.~Brooker, N.~C.~Tsamis and R.~P.~Woodard,
  Phys.\ Rev.\ D {\bf 94}, no. 4, 044020 (2016)
  doi:10.1103/PhysRevD.94.044020
  [arXiv:1605.02729 [gr-qc]].

\end{thebibliography}
\end{document}